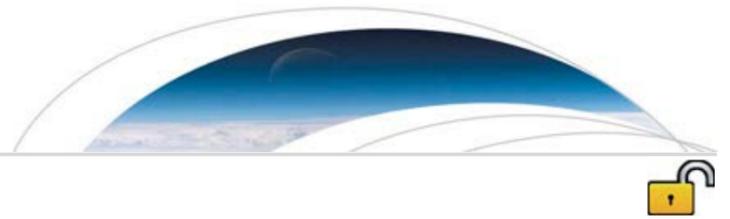

# The interaction between transpolar arcs and cusp spots

R. C. Fear[1], S. E. Milan[2,3], J. A. Carter[2], and R. Maggiolo[4]

[1]Department of Physics and Astronomy, University of Southampton, Southampton, UK, [2]Department of Physics and Astronomy, University of Leicester, Leicester, UK, [3]Birkeland Centre for Space Sciences, University of Bergen, Bergen, Norway, [4]Belgian Institute for Space Aeronomy, Brussels, Belgium

**Abstract** Transpolar arcs and cusp spots are both auroral phenomena which occur when the interplanetary magnetic field is northward. Transpolar arcs are associated with magnetic reconnection in the magnetotail, which closes magnetic flux and results in a "wedge" of closed flux which remains trapped, embedded in the magnetotail lobe. The cusp spot is an indicator of lobe reconnection at the high-latitude magnetopause; in its simplest case, lobe reconnection redistributes open flux without resulting in any net change in the open flux content of the magnetosphere. We present observations of the two phenomena interacting—i.e., a transpolar arc intersecting a cusp spot during part of its lifetime. The significance of this observation is that lobe reconnection can have the effect of opening closed magnetotail flux. We argue that such events should not be rare.

## 1. Introduction

During periods of northward interplanetary magnetic field (IMF), magnetospheric dynamics move to higher latitudes than those observed during periods of southward IMF. In this paper, we present simultaneous observations of two auroral phenomena associated with northward IMF: a transpolar arc and a cusp spot. We discuss the interaction between the two phenomena and the interesting magnetospheric topology that can be inferred.

Transpolar arcs extend from the nightside auroral oval into the polar cap. They were first observed on a global scale by *Frank et al.* [1982] but form part of a wider group of polar cap arcs, observations of which date back to the early Antarctic expeditions [*Mawson*, 1925]. It is well established that polar cap arcs occur during periods of low magnetic activity and hence northward IMF [*Davis*, 1963; *Berkey et al.*, 1976; *Gussenhoven*, 1982]. The cause of transpolar arcs has been widely debated [*Zhu et al.*, 1997; *Kullen et al.*, 2002; *Østgaard et al.*, 2003; *Milan et al.*, 2005; *Fear and Milan*, 2012a, and references therein]. However, recent results have indicated that the timescales on which the location of transpolar arcs depends on the IMF $B_y$ component and the presence of nightside ionospheric flows called "TRINNIs" [e.g., *Grocott et al.*, 2003, 2007] are strongly supportive of a formation mechanism for transpolar arcs based on the closure of lobe flux by magnetotail reconnection [*Milan et al.*, 2005; *Fear and Milan*, 2012a, 2012b; *Kullen et al.*, 2015]. In this mechanism, newly closed field lines which cross the equator near midnight have no clear return direction and so build up to form a wedge of closed flux in the magnetotail, the footprints of which map into the polar cap forming a transpolar arc [*Milan et al.*, 2005]. However, care must be taken to distinguish properly between transpolar arcs and other auroral phenomena which are more accurately described as poleward moving auroral forms and hence related to dayside processes [*Carter et al.*, 2015].

The cusp spot is the auroral signature of reconnection at the high-latitude (lobe) magnetopause [*Sandholt et al.*, 1996, 1998; *Øieroset et al.*, 1997; *Milan et al.*, 2000a]. High-latitude reconnection was first proposed by *Dungey* [1963]; in his original scenario, *Dungey* [1963] envisaged a closed magnetotail, and northward interplanetary magnetic field lines were assumed to reconnect simultaneously (or near simultaneously) in both hemispheres (dual lobe reconnection). *Russell* [1972] made two modifications to this picture: he incorporated an open magnetotail and suggested that it was more likely that a given interplanetary magnetic field line would reconnect in one hemisphere than both (single lobe reconnection). Subsequent studies considered alternative permutations [*McDiarmid et al.*, 1980; *Cowley*, 1981, 1983; *Reiff and Burch*, 1985] based on single or dual lobe reconnection with an open or closed magnetotail. Observational evidence has been reported for both single [*Gosling et al.*, 1991; *Kessel et al.*, 1996; *Milan et al.*, 2000b; *Twitty et al.*, 2004] and dual lobe reconnection [*Øieroset et al.*, 2005; *Lavraud et al.*, 2006; *Imber et al.*, 2006, 2007; *Slapak et al.*, 2015] from both spacecraft and ionospheric observations. Since the magnetotail is open (perhaps with rare exceptions [e.g., *Zhang et al.*, 2009]), the scenarios based on an open magnetotail have naturally received the most attention, and so it is generally understood that dual lobe reconnection has the effect of closing lobe flux [*Cowley*, 1981, 1983; *Imber et al.*, 2006, 2007], whereas single lobe reconnection causes no net change in the open flux content of the magnetosphere but simply "stirs" the open flux in the lobe [*Russell*, 1972; *Crooker*, 1992]. The brightness of the resulting cusp spot is modulated by solar wind density [*Frey et al.*, 2002; *Milan et al.*, 2010], and its local time is controlled by the IMF $B_Y$ component, consistent with the location of the lobe reconnection site being located in the region of highest magnetic shear [*Milan et al.*, 2000a]. Cusp spots may arise due to either single or dual lobe reconnection [*Imber et al.*, 2007]; simultaneous spots in the Northern and Southern Hemispheres may also arise due to the simultaneous occurrence of single lobe reconnection in both hemispheres [*Østgaard et al.*, 2005].

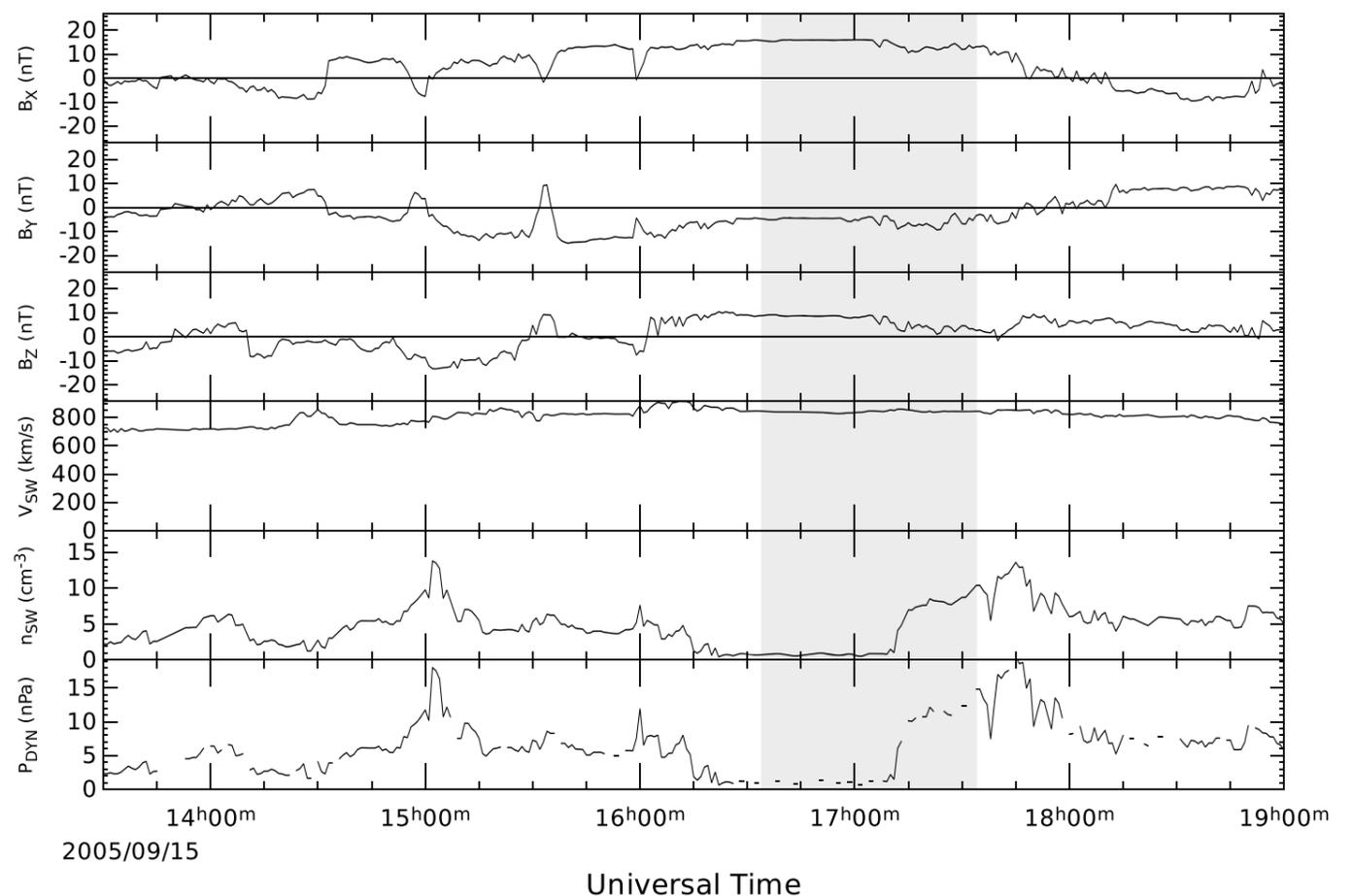

**Figure 1.** The solar wind conditions observed between 13:30 and 19:30 UT on 15 September 2005. From top, the GSM components of the IMF and the solar wind speed, density, and dynamic pressure. The gray box highlights the period during which a cusp spot was observed.

The magnetotail topology associated with transpolar arcs has been a topic of some controversy. The similarity between the plasma precipitation above transpolar arcs and that above the auroral oval is indicative that transpolar arcs are on closed field lines [*Frank et al.*, 1982, 1986; *Peterson and Shelley*, 1984; *Huang et al.*, 1989], though study of electrons above polar cap arcs have been used to argue for an open topology [*Hardy et al.*, 1982; *Gussenhoven and Mullen*, 1989]. Strong evidence for a closed topology is provided by the presence of a double loss cone, which has been reported at altitudes below 3.5 $R_E$, above transpolar arcs [*Menietti and Burch*, 1987]. *Fear et al.* [2014] recently presented observations of a transpolar arc observed by the IMAGE spacecraft on 15 September 2005 and simultaneous observations of the conjugate plasma environment observed deeper in the magnetotail by the Cluster spacecraft. Between 17:00 and 19:00 UT, Cluster intermittently observed a hot plasma population on closed field lines embedded in the open field lines of the high-latitude lobe, far from the nominal location of the plasma sheet. The authors argued that these observations, coupled with the complex motion of the arc (dawnward, then duskward, then dawnward again) provided strong evidence for the formation of transpolar arcs by magnetotail reconnection as proposed by *Milan et al.* [2005]. Since a cusp spot is a signature of lobe reconnection, which has been explicitly invoked to explain the motion of transpolar arcs [*Milan et al.*, 2005], we might expect cusp spots and transpolar arcs to be associated. In this paper, we present further observations of the same interval and show that a cusp spot was present for some of the lifetime of the transpolar arc. The arc was observed to move toward the cusp spot and intersect it. We argue that not only is this scenario likely to be common, but it indicates that the topologies associated with lobe reconnection are modified and single lobe reconnection then has the net effect of opening the closed flux which forms the arc.

## 2. Instrumentation

The auroral observations reported by *Fear et al.* [2014] were obtained by the far ultraviolet/wide-band imaging camera (FUV/WIC) camera on the IMAGE spacecraft [*Mende et al.*, 2000]. WIC is sensitive to emissions at wavelengths between 140 and 190 nm and provides global-scale auroral observations at a cadence of 2 min. At apogee, a pixel in a WIC image corresponds to a footprint of $52 \times 52$ km. During the interval covered in this paper, IMAGE was situated in the Southern Hemisphere observing the aurora australis.

Supporting observations of the interplanetary magnetic field and the solar wind density are provided. These data are provided by the high-resolution OMNI data set, which for this interval consists of data from the ACE spacecraft, lagged to the bow shock as described by *King and Papitashvili* [2005].

## 3. Observations

The lagged solar wind conditions observed between 13:30 and 19:30 UT on 15 September 2005 are summarized in Figure 1. The IMF observed by ACE turned northward at 16:00 UT and remained so until just after 19:00 UT. During this time, the IMF $B_Y$ component rotated from negative to positive. The solar wind speed remained steady at approximately 800 km s$^{-1}$, but the density varied dramatically throughout the interval, ranging from 0.4 to 13 cm$^{-3}$, resulting in significant variations in the dynamic pressure. During the period of interest, Geotail was situated just upstream of the bow shock, except for a crossing of the shock into the magnetosheath between 16:20 and 17:20 UT. While Geotail was in the solar wind, the magnetic field and plasma velocity, density, and dynamic pressure were an excellent match for the OMNI parameters (not shown), providing a high degree of confidence in the lag applied to ACE.

The global scale auroral observations are summarized in Figure 2. Each panel shows the IMAGE WIC observations at the time indicated, projected onto a magnetic latitude/magnetic local time grid. An animated version of all of the images underlying Figure 2 is included in Movie S1 in *Fear et al.* [2014]. The evolution and dynamics of the transpolar arc have been discussed by *Fear et al.* [2014], but in summary, a brightening occurs just poleward of the main oval at about 21 MLT and 16:10 UT (indicated by an arrow in Figure 2b), shortly after the northward turning of the IMF (Figure 1); this evolves into a feature which grows into the polar cap (Figures 2c–2e), forming a transpolar arc (shown by the arrow in Figure 2e). The transpolar arc persists during the period of northward IMF; after the IMF turned southward at 19:10 UT, the arc was "flushed out" of the polar cap (as described by *Milan et al.* [2005]) and disappeared after 20:10 UT (see Figure 4 and Movie S1 of *Fear et al.* [2014]).

An additional feature not discussed by *Fear et al.* [2014] is that a cusp spot was visible between 16:34 and 17:34 UT. The spot is indicated by a white arrow in Figures 2f–2n, and the period during which it was observed is highlighted by gray shading in Figure 1. Both the appearance and disappearance of the spot is rapid—it is not evident at 16:30 or 17:36 UT. Interestingly, the spot appeared while the solar wind density was low and disappeared just before the peak density was observed (Figure 1). However, the spot does appear to be well ordered by the IMF orientation; it appeared during the period of northward IMF and shortly after the $B_Y$ component had decreased in magnitude, reducing the clock angle. The spot disappeared just as the IMF $B_Z$ component decreased briefly toward zero, although it did not reappear (not shown) when $B_Z$ increased again at 17:45 UT, even though the $B_Y$ component then reversed, passing through zero, while the solar wind density peaked (Figure 1). This period corresponded to a reduction in the IMF $B_X$ component, which could be construed as making lobe reconnection in the Southern Hemisphere less likely. However, the absence of a cusp spot after 17:34 UT does not indicate that lobe reconnection ceased—the motion of the transpolar arc indicates that lobe reconnection continued [*Milan et al.*, 2005; *Fear et al.*, 2014].

The cusp spot appeared shortly before the transpolar arc. Between 16:38 and 17:00 UT, the arc varied in intensity, but its sunward end moved toward the spot (Figures 2e–2h). Then, from about 17:08 UT, the two intersected. This intersection persisted until the spot disappeared (Figures 2i–2n), though the transpolar arc was faint toward the end of this interval.

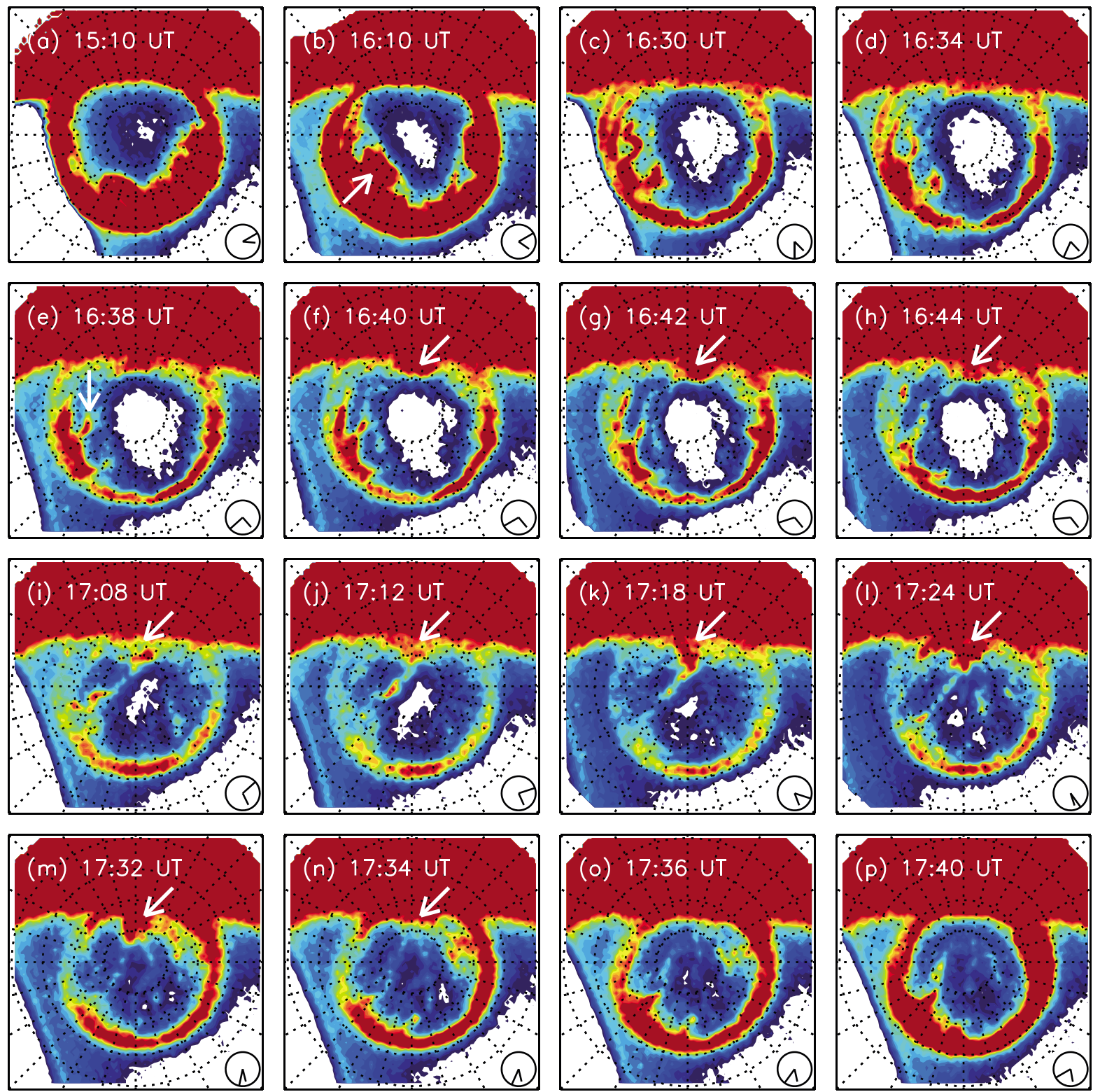

**Figure 2.** A montage of the Southern Hemisphere aurora observed by the IMAGE spacecraft. (a–p) The images plotted on a magnetic local time/magnetic latitude grid with noon MLT at the top of each panel, midnight at the bottom, dawn on the right, and dusk on the left; therefore, although the images pertain to the Southern Hemisphere, they are shown as if viewed from the Northern Hemisphere, through the planet. The white arrows highlight features discussed in the text.

## 4. Discussion

The presence of a transpolar arc indicates the presence of closed magnetic flux extending to high latitudes in the magnetotail lobes [*Frank et al.*, 1986; *Huang et al.*, 1987, 1989; *Milan et al.*, 2005; *Fear and Milan*, 2012a, 2012b; *Fear et al.*, 2014]. The fact that the transpolar arc connects the nightside and dayside ovals between 17:08 and approximately 17:34 UT (i.e., forms a full "theta") indicates that between these times, the magnetotail was totally closed in the azimuthal sector corresponding to the transpolar arc; in other words, the plasma sheet in that sector had expanded through magnetotail reconnection to be as thick as the entire magnetotail. This situation is sketched in Figure 3, which shows selected magnetic field lines either directly inferred from or modified from the *Tsyganenko* [1996] magnetic field line model. The black and gray lines are closed and open model field lines, respectively, in the dusk sector. (Model field lines are taken to be open if they do not close within 999 steps.) The blue lines are field lines nearer noon or midnight, which are closed in the

(a) View from the flank

(b) View from above the northern hemisphere

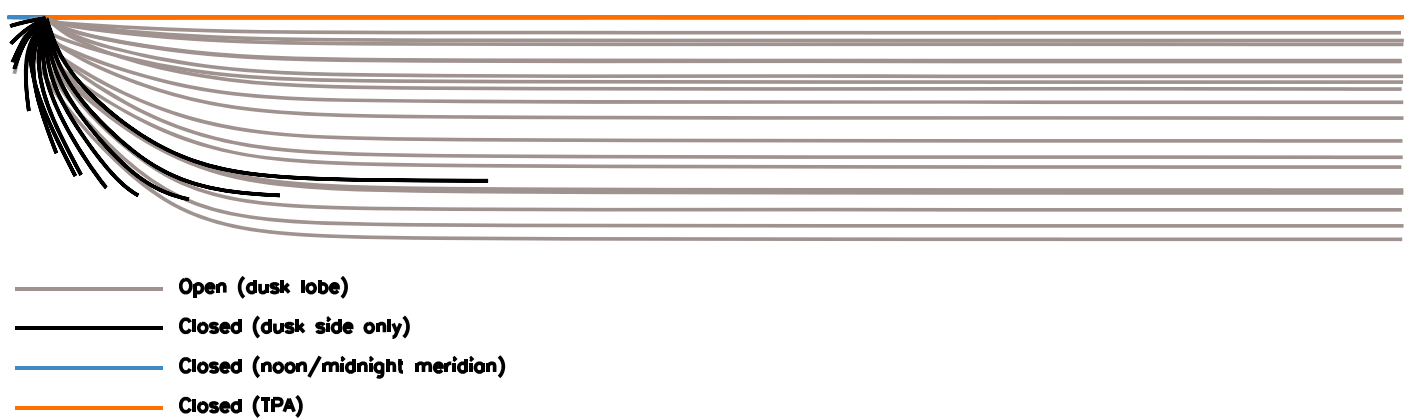

———— Open (dusk lobe)
———— Closed (dusk side only)
———— Closed (noon/midnight meridian)
———— Closed (TPA)

**Figure 3.** A sketch of the magnetotail configuration during a period when a transpolar arc fully extends across the polar cap (i.e., forms a full "theta"). Black and gray lines are Tsyganenko model magnetic field lines on the dusk side of the magnetosphere (closed and open, respectively). Blue and orange lines are magnetic field lines nearer the noon/midnight meridian. The blue field lines are closed in the Tsyganenko model; the orange field lines would normally be open (i.e., form part of the lobe), but in this narrow local time sector they are closed and map to the transpolar arc. In this case, the sector of closed (orange) field lines has been drawn in the midnight meridian, but in practice it will often be tilted and/or displaced from this location as described by *Milan et al.* [2005]. An animated version, rotating through different viewpoints, is provided in the supporting information.

*Tsyganenko* [1996] model (both the dayside and the nightside plasma sheet). The orange field lines are also in the midnight meridian; they correspond to open lobe magnetic field lines in the *Tsyganenko* [1996] model, but in this figure they have been modified to be closed in the magnetotail as proposed by *Milan et al.* [2005]. These orange closed field lines extend far downtail and map into the polar cap, with the precipitation thereon producing the transpolar arc.

Since it is common for a transpolar arc to extend fully across the polar cap, Figure 3 represents a common configuration of the magnetotail during periods when a transpolar arc is present. We note that there are some similarities between this sketch and Figure 3b of *Kullen* [2012], though the underlying interpretation differs—*Kullen* [2012] attributes the presence of high-latitude closed field lines directly to the propagation of a new plasma sheet twist direction downtail, rather than being built up by magnetotail reconnection.

A cusp spot indicates the occurrence of lobe reconnection; therefore, the intersection of the cusp spot with the transpolar arc indicates that at these times, the interplanetary magnetic field was undergoing reconnection with *closed* magnetic field lines in the magnetotail (which thread the transpolar arc). This scenario is sketched in a 2-D cut in Figure 4a. Consequently, the net effect of single lobe reconnection changes. Rather than simply redistributing open flux within the polar cap [*Russell*, 1972], reconnection opens the outer layer of closed flux which forms the transpolar arc (which maps to the sunward tip of the arc). The newly opened field lines then form part of the cusp spot. Further lobe reconnection with closed field lines will continue as adjacent closed field lines are drawn into the reconnection site, though this process may have a finite lifetime. Initially, the field lines that are drawn in to the lobe reconnection site will come from an azimuthal extent of the lobe that is only slightly wider than the extent of the reconnection line (Figure 4b). However, as shown, the lobe flows converge on the reconnection site indicating that as lobe reconnection continues, the field lines drawn in to the reconnection site originate from gradually wider azimuthal regions (at lower latitudes). If the spatial extent of closed field lines associated with the transpolar arc is reasonably uniform across latitudes, the proportion of reconnecting field lines that are open beforehand will gradually increase. This suggests that at some point, the net opening of transpolar arc field lines may cease, to be replaced by regular "stirring" again, though further observations will be necessary to test this speculation.

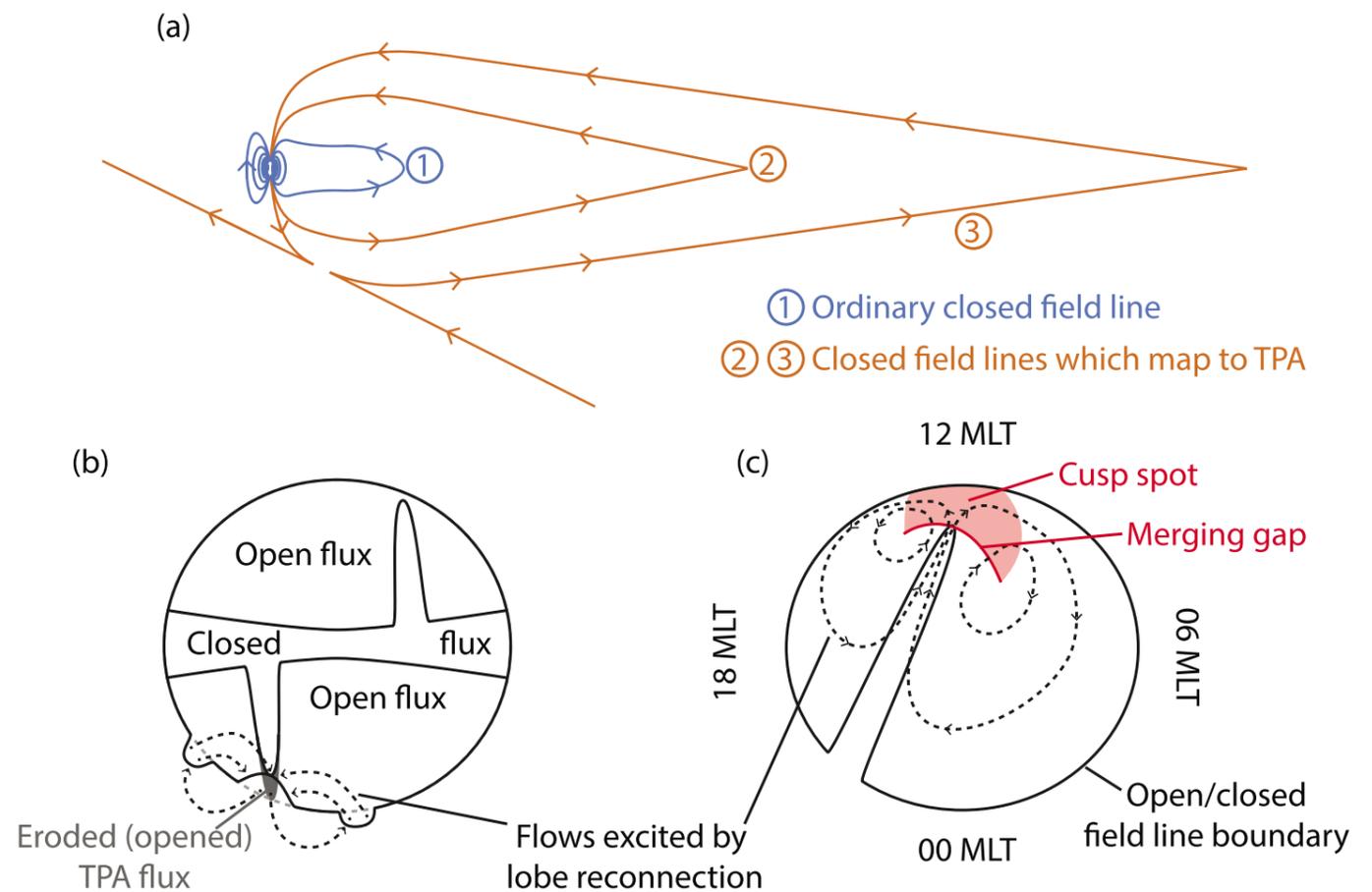

**Figure 4.** A sketch illustrating the consequences of lobe reconnection when a transpolar arc intersects a cusp spot. (a) The field lines in the local time sector of the "wedge" of closed flux in the magnetotail which maps to the transpolar arc, following the same color scheme as in Figure 3. Field line ③ is shown undergoing single lobe reconnection with an interplanetary magnetic field line, which results in a change of topology as the transpolar arc field line is opened. (b) Cross section of the magnetotail, looking toward Earth, following the same format as Figure 7a in *Milan et al.* [2005]. Closed field lines are shown sandwiched between the open lobes, with a tongue of closed flux forming the transpolar arc (TPA). Lobe flows and the erosion of the TPA are indicated. (c) The polar cap dynamics corresponding to the opening of transpolar arc magnetic flux by single lobe reconnection, shown in the same format as Figure 2. The adiaroic open/closed field line boundary is perturbed significantly into the polar cap, indicating the location of the transpolar arc. At the point where the transpolar arc intersects the merging gap, magnetic flux is opened and transported into the lobe proper. Field lines crossing the merging gap either side of the transpolar arc (which are therefore open) undergo reconnection without a change in topology.

The scenario in Figure 4a shares some similarities with that proposed by *McDiarmid et al.* [1980, Figure 9] and *Cowley* [1981, 1983, Figure 8d in both papers]. However, it differs in that the magnetotail is not closed in all local times. If the lobe reconnection site is wider than the extent of the closed "wedge" of magnetotail flux, then the lobe reconnection process will also occur with ordinary open lobe field lines on either side. Therefore, either side of the transpolar arc, there will be no change in the topology of field lines which will simply be "stirred" [*Crooker*, 1992].

The situation in the ionosphere is sketched in Figure 4c. The polar cap is sketched with noon MLT at the top and dawn to the right. The solid black line indicates the open-closed boundary, perturbed into the polar cap by the presence of the transpolar arc [cf. *Milan et al.*, 2005, Figures 8 and 9; *Fear and Milan*, 2012a, Figure 14]. The red line indicates the merging gap, which is the ionospheric projection of the lobe reconnection site. The cusp spot (shaded pink) forms on field lines which have recently undergone reconnection and hence lies on open field lines immediately downstream of the merging gap; therefore, the merging gap lies at the poleward edge of the cusp spot. Magnetotail flux will be drawn in to the location of the cusp spot, where it undergoes reconnection and is then transferred from the local time sector of the reconnection site to earlier/later local times. The restoration of pressure balance with the solar wind results in antisunward flows on either side [*Cowley and Lockwood*, 1992]; the resulting lobe convection cells are indicated by the dashed black lines. Contrary to the standard single lobe reconnection scenario, a topology change (from closed to open field lines) occurs at the intersection between the merging gap and the transpolar arc.

Since the transpolar arc is the auroral signature of closed field lines which are trapped in the lobe, the field lines mapping through the arc simply move with the surrounding open field lines [*Milan et al.*, 2005; *Goudarzi et al.*, 2008]. The ionospheric flows excited by lobe reconnection will feed in to the location of the cusp spot. Consequently, if a transpolar arc exists but does not initially intersect the cusp spot, the arc will move toward the spot [see *Milan et al.*, 2005, Figures 8 and 9; *Fear and Milan*, 2012a, Figure 14]. As long as lobe reconnection persists long enough, the transpolar arc will ultimately end up pointing toward and intersecting the cusp spot, and so the situation observed in Figure 2 and the opening of closed magnetotail flux sketched in Figure 4 will be relatively common as a proportion of periods with transpolar arcs.

As transpolar arcs are a northward IMF phenomenon, they are associated with the suppression of low-latitude reconnection. They are also observationally associated with the occurrence of magnetotail reconnection [*Milan et al.*, 2005; *Fear and Milan*, 2012b]; therefore, one would typically expect a gradual decrease in the area of the polar cap during the lifetime of the transpolar arc [*Milan et al.*, 2005]. If one considers the polar cap to include the transpolar arc (rather than to refer strictly to the region of open field lines), then the decrease in area is due to the closure of field lines on either side of the field lines which form the "wedge" of closed flux mapping to the arc, since these field lines succeed in convecting back to the dayside as proposed by *Dungey* [1961]. If the rate of magnetotail reconnection is small and/or the extent of the magnetotail reconnection line is not much wider than the region in which the "wedge" forms, then the rate at which the polar cap area decreases will be small (possibly negligible). The occurrence of lobe reconnection with closed magnetotail field lines mapping through the transpolar arc will result in a change in the proportion of closed flux in the magnetotail. However, this process will not of itself cause the main auroral oval to contract, since the latitude of the poleward edge of the oval is determined by the combined flux content of the (open) lobe and the (closed) transpolar arc. Lobe reconnection may transfer flux between these two categories but will not of itself change the total of these two fluxes. Indeed, visual inspection of Figure 2 indicates that there was no significant change in the area of the polar cap over the lifetime of the cusp spot.

In the above, we have considered the more common scenario of single lobe reconnection. However, more complicated scenarios potentially exist should dual lobe reconnection occur while a transpolar arc is present. Here one would expect a transfer of closed flux from the nightside (the transpolar arc) to the dayside; there will be no overall net change in topology, although field lines will be briefly opened then reclosed (on the dayside) if reconnection in the two hemispheres is not exactly simultaneous.

## 5. Conclusions

We have presented observations of the intersection between a transpolar arc and a cusp spot. When such an intersection occurs, it indicates that the single lobe reconnection process is modified such that it opens some of the closed magnetotail flux which maps through the transpolar arc, rather than the standard scenario in which it simply "stirs" open magnetotail flux. Since transpolar arcs move with the surrounding (open) lobe flux, the opening of transpolar arc magnetic field lines by lobe reconnection should be reasonably common, as a transpolar arc at a different local time will be drawn toward the cusp spot. This will particularly be common in cases where transpolar arcs are observed to move, since transpolar arc motion indicates the presence of ongoing lobe reconnection.

**Acknowledgments**
R.C.F. was supported by the UK's Science and Technology Facilities Council (STFC) Ernest Rutherford Fellowship ST/K004298/1. S.E.M. and J.A.C. were supported by STFC Consolidated grant ST/K001000/1 and R.M. by the Belgian Science Policy Office through the Solar-Terrestrial Center of Excellence. We gratefully acknowledge support by the International Space Science Institute through funding of its International Team on Polar Cap Arcs. The IMAGE FUV data were provided by the NASA Space Science Data Center (NSSDC), and we gratefully acknowledge the PI of FUV, S. B. Mende of the University of California at Berkeley. The OMNI IMF data were obtained through NASA's CDAWeb (http://cdaweb.gsfc.nasa.gov/), for which we acknowledge J. H. King, N. Papatashvilli, and the principal investigators of the magnetic field and plasma instruments Advanced Composition Explorer (ACE) spacecraft.